\documentclass [12pt]{article}
\usepackage{amssymb,amsmath,amsthm}
\usepackage{graphicx}
\usepackage[T1]{fontenc}

\begin{document}

\title{Pairwise Comparisons Rating Scale Paradox}

\author{W.W. Koczkodaj \footnote{Partially supported Euro grant Human Capital, wkoczkodaj@cs.laurentian.ca}\\
 Computer Science, Laurentian University,  Sudbury, Ontario, Canada}
 
\date{2015}
\maketitle
\begin{abstract}
This study demonstrates that incorrect data are entered into a pairwise comparisons matrix for processing into weights for the data collected by a rating scale. 
Unprocessed rating scale data lead to a paradox. A solution to it, based on normalization, is proposed. This is an essential correction for virtually all pairwise comparisons methods using rating scales. The illustration of the relative error currently, taking place, is discussed. \\

\end{abstract}

\section{Introduction}

Thurstone's Law of Comparative Judgments, introduced \cite{Thurstone1927} in 1927 was a milestone in pairwise comparisons (PCs) research although the first documented use of PCs is traced to Ramond Llull in 13th century. A considerable number of customizations, based on different rating scales, have been proposed. Some of them have generated controversies which are not the subject of this study. This study is independent of pairwise comparisons customizations. It concentrates on the theoretical aspects of the rating scale, leaving to the originators of PCs customizations, to accommodate appropriate corrections. The starting point for this study is a PC matrix. 
Numerous examples demonstrated that the pairwise comparisons can be used to draw conclusions in a comparatively easy and elegant way. The brilliance of the pairwise comparisons could be reduced to a common sense rule: take two at a time if we are unable to handle more than that. For relating one item to another item in a pair, PCs relies on a rating scale ``1 to $m$'', where 1 denotes equality and $m$ is used to reflect superiority (``advantage'' or some kind of ``perfection'') of one item above the other item.

In simple language, a rating scale is a set of categories designed to elicit data about a quantitative or a qualitative stimuli (or attribute). It requires a rater to choose a numeric value, sometimes by using a graphical object (e.g., line), to the rated entity, as a measure of some rated stimuli.
One of the best known examples of such a scale is a ``scale of 1 to 10'', or ``scale from 1 to 10'' where 10 stands for some kind of perfection.


Graphical scales ``one to four'' or ``one to five'' are often represented by stars, especially when used on the Internet, for rating movies, services, etc. In colloquial English, the idiomatic adjective ``five-star'' (meaning ``first-rate'') is frequently used. The choice of the rating scale upper limit as 5 may have something do with the number of fingers in one hand. The use of $m=10$ may be influenced by the numerical system with 10 as the numerical base, which in turn is derived from the number of fingers on two hands. Rating scales can also include scores in between integers (or use graphics, such as a line to mark the answer with a vertical bar, $\times$, or another symbol) to give a more precise rating. The origin of rating scales are not clear but \cite{DA1978} seems to be one of the most cited (by, for example, the Web of Science count) and celebrated interpretations.

Pairwise comparisons have great application to collective intelligence since this method allows us to synthesize often highly subjective assessments of expert panels, steering committees, or other collective decision making constituencies. 
In case of doubt, it was evidenced in one of the flagship ACM publications \cite{FHH2010}.
 
Finally, this study does not invalidate two major contributions in \cite{DXLD2008} regarding the scale selection and \cite{CW2010} regarding the rating scale unit used for pairwise comparisons. In fact, both studies and their contributors have a considerable impact on this study. 

\section{The rating scale paradox}

Paradoxes serve a very important purpose in science. They stimulate creative thinking. Banach Tarski paradox is one of the most stunning in mathematics. Russell's paradox contributed to a drastic paradigm shift in the foundation of set theory. This paradox calls for data entry correction. In the current situation, the relative error for the scale 1 to 9 (in Section~\ref{example}) is 23\% for the value 2 which may be frequently entered since such a scale (with its own drawback) promotes the use of low and high values.\\

\noindent According to \cite{DA1995}:
\begin{quote}
Graded responses are used where there is no measuring instrument
of the kind found in the physical sciences, but the structure of the responses mirrors physical measurement. In physical measurement, the amount of a property of an entity
is measured by using an instrument to map the property onto a continuum which has been divided into units of equal length, and then the count o f the number o f units from the origin that the property covers (often termed simply the measurement of the entity), is the location of the entity with respect to that property. Although it is recognized that instruments have operating ranges, the measurements of any entity are not taken to be a function of the operating range of an instrument--if the property exceeds the range of one instrument, then an instrument with a relevant range is sought. In deterministic theories, the variations and sizes of the property measured are considered sufficiently large relative to the size of the unit, that errors of measurement are ignored, and the count is taken immediately as the measurement.
\end{quote}

Using rating scales for the data entry is a popular method in most pairwise comparisons methods to collect graded responses or assessments. However, they suffer from the paradox and the acquired data cannot be entered into a PC matrix without a prior processing. \\

We need to clarify the terminology. Ratios of entities (sometimes they are referred to as ``ratio scale values'') create a PC matrix. The ratio scale is not the same the rating scale. Various rating scales are used for acquiring ratios but not all ratios are taken from a ratio scale.  

We assume that $M$ is a reciprocal PC matrix over $\mathbb{R}^+$. Let $M$ be of the form:

\begin{displaymath}
M = \begin{bmatrix}
1 & m_{1,2} & \cdots & m_{1,n} \\ 
\frac{1}{m_{1,2}} & 1 & \cdots & m_{2,n} \\ 
\vdots & \vdots & \vdots & \vdots \\ 
\frac{1}{m_{1,n}} & \frac{1}{m_{2,n}} & \cdots & 1
\end{bmatrix}
\end{displaymath} \\

\noindent The following simple example of a PC matrix for three entities $A$, $B$, and $C$:

\begin{displaymath}
M = \begin{bmatrix}
    1     & 2     & 2 \\
    0.5   & 1     & 1 \\
    0.5   & 1     & 1 \\
\end{bmatrix}
\label{M221}
\end{displaymath}

\noindent reflects $A=2*C$, $A=2*B$, and $B=C$ hence $A=2, B=1, C=1$ is (one of many) solutions.

So far, there seems to be no problem with this PC matrix since the above PC matrix $M$ is consistent as the only triad in $M$ fulfills the consistency condition: 

\begin{equation}\label{cc} m_{ij} \cdot m_{jk}=m_{ik}\end{equation}
for every $i,j,k=1,2, \ldots ,n$. \\


There is one thing drastically missing in $M$: the rating scale upper limit. In other words, nothing will change if we use a different rating scale, say 1 to 3 (recommended in \cite{FKS2010}). If so, we may try a bit bigger rating scale upper limit: 1 to 101 (giving 100 slots). For such a scale, things get a bit complicated as such a rating scale makes $A=B=C$ in practical terms, since 2 comes so close to 1 that it is hard to see it ``with the naked eye'' as we can imagine increasing the rating scale upper limit to infinity. Section~\ref{example} shows that even a scale, of a moderate size 1 to 9, causes a substantial approximation error. 
Regardless of the practicality, the rating scale values 1 and 2 become practically indistinguishable for large $m$. Let us recall that 1 on the rating scale stands for equality of compared entities. Using 2 for ``two times bigger'' (or somehow ``superior'') on even a moderate rating scale ``1 to 10'' 
has never been considered as incorrect yet weights (computed by any method since $M$ is consistent)are: 

$$[0.5, 0.25, 0.25]$$ 

\noindent It reflects the fact $A=2*B=2*C$, although $A=B=C$ with as high accuracy as we can wish to have for $m \rightarrow \infty$. \\

\noindent Evidently:
$$\begin{array}{lcl} 
A=2*B &  &  \\ A=B 
\end{array}$$

\noindent give $A*1=A*2$ hence $1=2$ for A>0 with as high accuracy as we wish to have.

This is a pairwise comparisons rating scale paradox of a fundamental importance since we provided evidence that: $A=2*B$ and $A=B$ for $A>0$ and $B>0$.

The paradox takes place since the entries in the PC matrix 
do not have any connection to the rating scale upper limit. The value 2 on the rating scale ``1 to 10'' is not the same 2 as on the scale 1 to 101.
Evidently, the middle rating scale value depends on $m$ and for $m=101$, it is not 2 but 50.

\section{Solution to the paradox}

\begin{figure}[p]
\centering
\includegraphics[width=.75\linewidth]{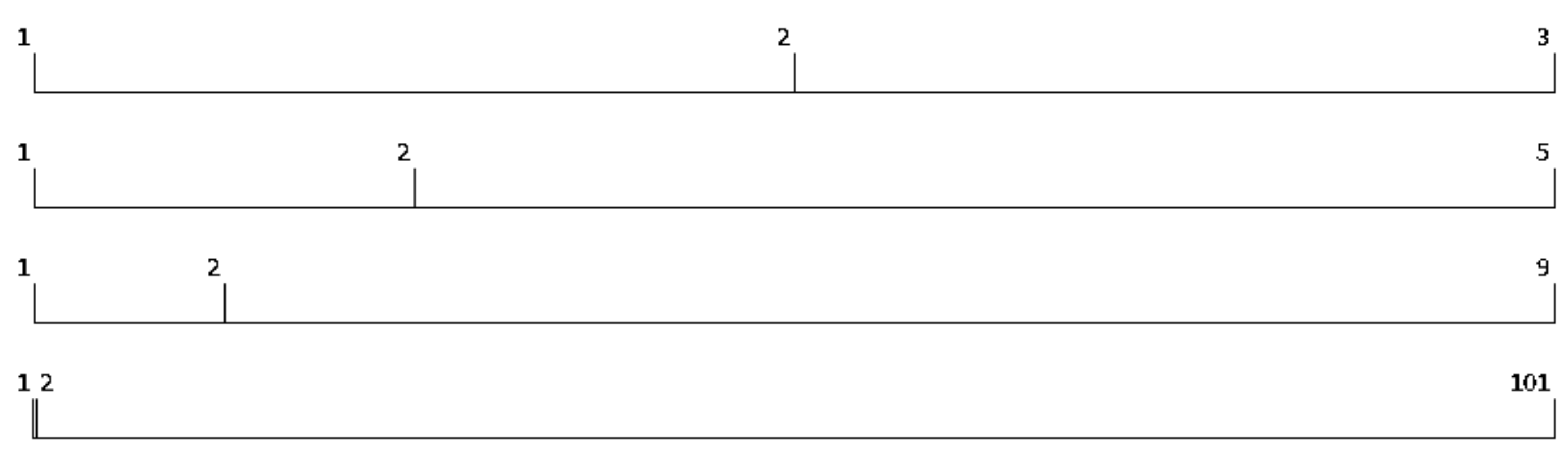}
\caption[scale9]{Value 2 on various rating scales}
\label{fig:slider}
\end{figure}

Let us look at Fig.~\ref{fig:slider}, representing value 2 on rating scales with the different upper limits. For $m=101$, it is not ``half of the rating scale'' . For $m=9$, ``the half'' is 4. With the increased rating scale upper limit $m$, the meaning of the value 2 shifts towards ``equality'' with the relative error diminishing to 0. For $v \rightarrow \infty$ on a rating scale with $m \rightarrow \infty$, the value to be entered into the PC matrix is 2 since 1 is the ``neutral'' point.

We need to incorporate $m$ into the PC matrix $M$.
A normalizing mapping is proposed as a solution. Technically, the term
\textit{normalization} should not be used since we are not mapping the rating scale values
to the interval $[0,1]$ (open or close). In a PC matrix, reflecting the equality of all entities is done by all entries having a value of 1. Evidently, 0 (as a ratio) does not exist.
The normalization of the ratio scale values into $[0, 1]$ interval and adding the ``neutral'' 1 prevents the paradox from taking place. For it, a linear function $$f:[1, m]\to [1,2]$$ such that $f(1)=1, f(m)=2$ needs to be defined (improved by \cite{EW}). It is given by:

$$f(x)=\frac{1}{m-1}x+\frac{m-2}{m-1}$$

Evidently, for $v\in [1, m]$, $ f(v)=\frac{v+m-2}{m-1}$  hence $ f(v)=1+\frac{v-1}{m-1}$.
For a given value $v$ on a scale ``1 to $m$'', the PC matrix entry should be:

\begin{equation}
 1+(v-1)/(m-1) 
\end{equation}

\noindent It is easy to see that: 

$$\lim_{m\to\infty} {1+(v-1)/(m-1)}=1$$ 

It means that our  paradox no longer takes place. Certainly, other solutions may be considered and the future research is expected to contribute to it. The research of new ``normalization'' methods is in progress. \\

\noindent \textbf{Example:} \\

The scale 1 to 6 is used in elementary schools in at least one of the EU countries. It has an interpretation for 2 as a \textit{marginally passing mark}. It has a definitely different meaning than 2 on the scale 1 to 101 as a hypothetical scale for evaluating University students.
In fact, assuming that 1 is the lowest score and 101 is the top score, it is hard to envision any school in any country setting 2 as satisfactory
score.

Certainly, the rating scale upper limit of 101 can be extended to any arbitrarily large value bringing 2 as close to 1 as we can imagine. It does not matter whether or not we use such a scale since a substantial error occurs (23\%) for even a relatively modest scale of 1 to 9 as evidenced by a numerical example in Section~\ref{example}.\\
$\blacksquare$

It is also worth noticing that our normalizing mapping transforms rating scale values into $[1, 2]$ (and their inverses $ [1/2, 1] $). It also creates PC matrices which have mathematically ``nice'' entries since their values are less than the F\"ul\"op constant (approx. 3:330191) exploited in \cite{FKS2010} to analyze the rating scale. The exact value of F\"ul\"op's constant is equals to:

\begin{equation}
a_0=((123+55\sqrt 5)/2)^{1/4}=\sqrt{\frac{1}{2} \left(11+5
\sqrt{5}\right)}\approx 3.330191
\end{equation}
Thanks to F\"ul\"op's constant, the optimization problem for finding weights can be transformed into the convex programming problem with a strictly convex objective function to be minimized (see \cite{Fulop08}, Proposition $2$):

\begin{equation}
\begin{array}{lll}
\min &&\sum\limits_{i=1}^{n-1} f_{a_{in}}(x_i)+
\sum\limits_{i=1}^{n-2}
\sum\limits_{j=i+1}^{n-1} f_{a_{ij}}(x_{ij})\\
{\rm s.t.} &&x_i-x_j-x_{ij}=0,\ i=1,\dots ,n-2,\ j=i+1,\dots ,n-1.
\end{array}
\label{eq:b8}
\end{equation}

\noindent where the univariate function is defined as:
\begin{equation} f_a(x)=\left (
e^x-a\right )^2+\left( e^{-x}-1/a\right )^2 \label{eq:b7}
\end{equation}

\noindent and $a$ is replaced by the F\"ul\"op's constant. 

\noindent

%
%
%
%
%

\section{A numerical example}
\label{example}

\begin{figure}[h]
\centering
\includegraphics[width=1\linewidth]{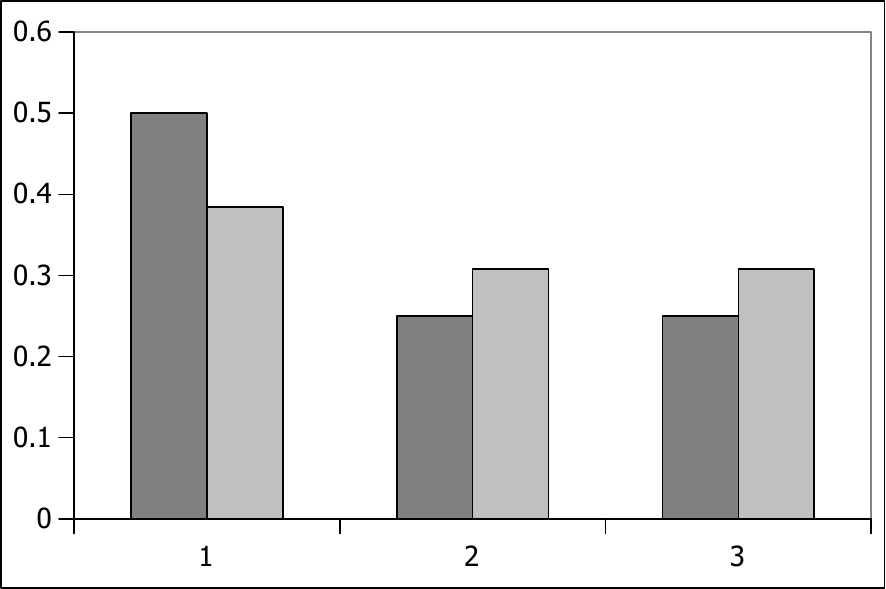}
\caption[Weights for the original and corrected data]{Weights for the original and corrected data}
\label{fig:2cases}
\end{figure}

For a rating scale 1 to 9 
value 2 gives what Fig.~\ref{fig:2cases} demonstrates. The original and corrected values for this scale are in Table~\ref{tab:corr9}. 
\begin{table}[h]
  \centering
  \caption{The original and corrected values for the scale 1 to 9}
    \begin{tabular}{rrrrrrrrr} 
   \hline 
    1     & 2     & 3     & 4     & 5     & 6     & 7     & 8     & 9 \\
    1     & 1.125 & 1.25  & 1.375 & 1.5   & 1.625 & 1.75  & 1.875 & 2 \\
  \hline
    \end{tabular}%
  \label{tab:corr9}%
\end{table}%

Two $3 \times 3$ PC matrices  with the original input data and the adjusted data by the geometric means (unnormalized and normalized to 1) are illustrated by Fig.~\ref{fig:2cases}. Bars on the left demonstrate the original data. Bars on the right are for the normalized data. 

\begin{figure}[p]
\centering
\includegraphics[width=0.4\linewidth]{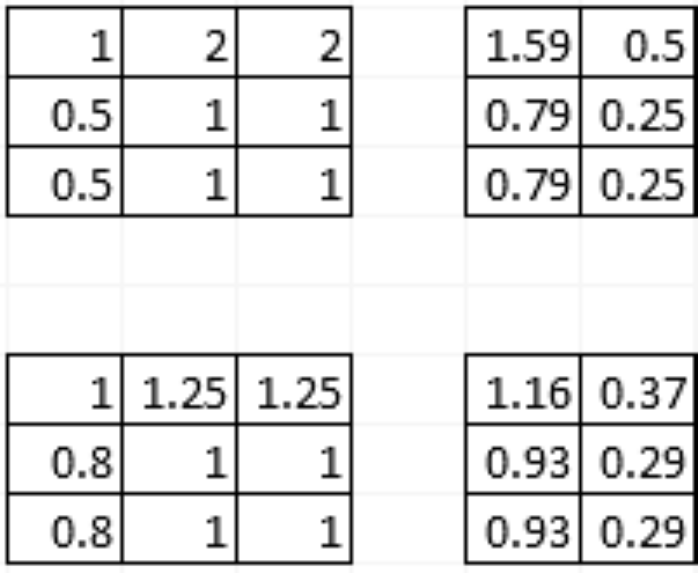}
\caption[PC matrices with the original input data and the adjusted data]{PC matrices with the original input data and the adjusted data}
\label{fig:2casesNum}
\end{figure}

\noindent The relative error for the above results:
$$
    \eta = \frac{\epsilon}{|v|} = \left| \frac{v-v_\text{approx}}{v} \right| = \left| 1 - \frac{v_\text{approx}}{v} \right|, 
$$

\noindent is: 23\% for all three entities which can hardly be ignored. Pretending that ``nothing happened'' is not an option when such a scale is used for a project of national importance (e.g., safety of a nuclear power station).\\

\section{Input data without the paradox effect}

Evidently, rating scale entries cannot be directly entered into a PC matrix without prior processing. The proposed  normalization prevents the paradox from taking place. In particular, experiments with randomly generated bars in \cite{Kocz1996, Kocz1998} render not only the correct results but evidence that the estimation error decreases when pairwise comparisons are used.

The Stone Age, mentioned in this author's former publications, and the ratio of stone weights give also the correct entries for the direct inclusion into a PC matrix since no rating scale is involved in it. The same goes for other physical measurements (distance, time) which include 0. However, there is a problem with the temperature expressed in Celsius scale as 0 is not the absolute 0 as in Kelvin (correct) scale.

It is also important to point out that most fuzzy extensions of PCs (of which the most cited in Web of Science is \cite{VP1983}) do not suffer from the presented paradox since they have a membership function with values in $[0,1]$. However, it also requires further scientific examination.

\section{Conclusions}

Entering categorical data into a PC matrix, without prior preprocessing, leads to a paradox. The absence of the rating scale upper limit in a PC matrix is the result for such a paradox. The willpower of inventors of various pairwise comparisons customizations   cannot change this situation. ``Having in mind'' the rating scale upper limit, without incorporating it into processing, cannot help. Data acquired by a rating scale must be processed before they are entered into a PC matrix for weights.
It does not matter what processing we use for the original rating scale entries to obtain weights (for example, heuristic methods specified in \cite{GM} or \cite{KK}) when we fail to incorporate the rating scale upper limit into the PC matrix. In other words, we cannot increase the number of aces in the deck of cards by shuffling them. It is imperative for the input data acquired by a rating scale to be normalized. 

Fig.~\ref{fig:slider} adequately illustrates that the meaning of the value 2 on a rating scale depends on the upper limit of the rating scale. Evidently, for a rating scale 1 to $m$ with $m \rightarrow \infty$, the value 2 is shifted towards 1 not only visually but these two values become indistinguishable as their difference (2-1=1) on this scale with the length $m-1$ becomes infinitely small. For a large $m$ (say, $10^{999}$), it does not much matter if we select 1 or 2 since these are both small on such a scale.

The pairwise comparisons method is the most amazing and the universal approach to assessments and decision making problems. Even for the incorrect entries, the received results were remarkably useful. Redoing computations for former projects with the proposed correction may contribute to providing a better evidence of the superiority of the pairwise comparisons method in terms of higher precision. The bad news is that most former publications on pairwise comparisons should be redone if they used data acquired by a rating scale unless they preprocessed input data by a method which prevents the paradox from occurring. The good news for authors of such studies and University administrators is that they may improve their publication record.

\section*{Acknowledgments}
This project has been supported in part by the Euro Research grant ``Human Capital''. 
The author is grateful to Grant O. Duncan (Team Lead, Business Intelligence and Software Integration, Health Sciences North, Sudbury, Ontario) for his help with proofreading this text. Matteo Brunelli and William C. Wedley (listed in the alphabetical order) have commented on the first draft and inspired many enhancements of this presentation. For it, the author is very grateful to them. The author also acknowledges that improvements and extensions to my original mathematical formulas (\cite{EW}), done by Eliza Wajch, are of great importance. Numerous researchers on four continents (Australia, Asia, Europe, and North America) have been extremely supportive through this project and I would like to thank all of them.

\end{document}